# Intensity Discriminability of Electrocutaneous and Intraneural Stimulation Pulse Frequency in Intact Individuals and Amputees

Jacob A. George, *Student Member, IEEE,* Mark R. Brinton, Paul C. Colgan, Garrison K. Colvin, Sliman J. Bensmaia, and Gregory A. Clark

*Abstract*— Electrical stimulation of residual nerves can be used to provide amputees with intuitive sensory feedback. An important aspect of this artificial sensory feedback is the ability to convey the magnitude of tactile stimuli. Using classical psychophysical methods, we quantified the just-noticeable differences for electrocutaneous stimulation pulse frequency in both intact participants and one transradial amputee. For the transradial amputee, we also quantified the just-noticeable difference of intraneural microstimulation pulse frequency via chronically implanted Utah Slanted Electrode Arrays. We demonstrate that intensity discrimination is similar across conditions: intraneural microstimulation of the residual nerves, electrocutaneous stimulation of the reinnervated skin on the residual limb, and electrocutaneous stimulation of intact hands. We also show that intensity discrimination performance is significantly better at lower pulse frequencies than at higher ones – a finding that's unique to electrocutaneous and intraneural stimulation and suggests that supplemental sensory cues may be present at lower pulse frequencies. These results can help guide the implementation of artificial sensory feedback for sensorized bionic arms.

## I. INTRODUCTION

In the United States alone, nearly one in every 200 individuals suffers from limb-loss [1]. The current standard-of-care for upper-limb amputees is unsatisfactory, and as a result, up to 50% of upper-limb amputees abandon their prostheses [2], often citing a lack of sensory feedback as a primary reason [3].

Recent work has shown that conveying sensory feedback through electrical stimulation of the peripheral nervous system yields functional and psychological benefits [4]–[14]. A key component of this is the ability to confer the magnitude of a tactile stimulus, thereby enabling the fine dexterity necessary for fragile object manipulation [13], [15], [16].

To this end, we applied classical psychophysical methods to measure participants' ability to discriminate changes in electrical stimulation pulse frequency. This information can be used to estimate the maximum number of distinguishable gradations that could be conveyed by a sensorized bionic arm. We explored this question from two extremes: 1) using highly invasive and highly specific intraneural microstimulation, and 2) using non-invasive, non-specific electrocutaneous stimulation. Our results can be used guide the implementation of artificial sensory feedback for sensorized bionic arms.

## II. METHODS

### A. Human Subjects

A total of eight human participants were recruited for this study: seven intact subjects and one amputee participant who had a transradial amputation 13 years prior to the present experiments. The four female and four male participants were between the ages of 20 and 60. Additional information regarding the participants is located in Table 1.

Informed consent and experimental protocols were carried out in accordance with the University of Utah Institutional Review Board.

### B. Stimulation Devices

The amputee participant had two, 100-electrode Utah Slanted Electrode Arrays (USEAs; Blackrock Microsystems, Salt Lake City, UT, USA) implanted into his residual nerves; one in the median nerve and one in the ulnar nerve. Additional information regarding the devices and implantation procedure used for this participant can be found in [12], [13], [17].

The intact participants and amputee participant (2.5 years after the USEAs were explanted) received electrocutaneous stimulation through a custom-fabricated silicon stimulation pad. The stimulation pad was a 9-cm$^2$ square pad, 4-mm thick, and consisted of one 0.79-cm$^2$ (1 cm diameter) stimulating electrode surrounded by four 0.44-cm$^2$ (0.75 cm diameter) ground electrodes.

Electrical stimulation was delivered using the Grapevine System with Micro2+Stim (intraneural) and Micro+Stim (electrocutaneous) front ends (Ripple Neuro LLC, Salt Lake City, UT, USEA). The electrocutaneous signal was amplified with a custom amplifier (-150V/+150V compliance, modified from [18]) prior to delivery.

### C. Stimulation Parameters

Intraneural stimulation was delivered simultaneously across ten electrodes on the median USEA to evoke a percept on the palmar side of the hand, between the index finger and thumb. Stimulation was delivered as biphasic, cathodic-first pulses, with 300–320-µs phase durations, and a 100-µs interphase duration. The pulse frequency varied between 12.5–167 Hz, and stimulation amplitude was fixed at 90 µA.

For intact participants, electrocutaneous stimulation was delivered through a single electrode placed on the base of the palm between the thumb and index finger. For the amputee, electrocutaneous stimulation was positioned on the residual limb such that the evoked percept was felt on the phantom palm between the index finger and thumb. Electrocutaneous stimulation consisted of biphasic, cathodic-first pulses, with 100 µs phase durations, and a 100- µs interphase duration. The pulse frequency varied between 12.5 and 200 Hz. Due to variations in skin impedance, the stimulation amplitude was chosen individually for each participant. Stimulation amplitude was typically set 1.5 mA above the threshold of detection; however, in the amputee, we increased amplitude to 12.2 mA (the maximum allowed by the stimulator and about 6 mA above threshold) because the participant reported the percept as too faint and inconsistent at lower amplitudes (Table 1).

Table 1. Electrocutaneous stimulation for intact and amputee participants

| Age | Gender | Amputation | BMI (kg/m$^2$) | Stim Location | Impedance (kΩ) | Stim Threshold (mA) | Stim Amplitude (mA) |
|---|---|---|---|---|---|---|---|
| 20 | Male | N/A | 29.8 | Left palm | 24.55 | 1.3 | 2.8 |
| 30 | Female | N/A | 24.1 | Left palm | 15.6 | 1.2 | 2.7 |
| 27 | Male | N/A | 23.1 | Left palm | 18.48 | 1.5 | 3.0 |
| 24 | Female | N/A | 25.9 | Left palm | 14.53 | 1.6 | 2.9 |
| 21 | Female | N/A | 22.3 | Left palm | 16.8 | 1.8 | 3.3 |
| 33 | Female | N/A | 20.7 | Left palm | 20.55 | 1.6 | 3.1 |
| 36 | Male | N/A | 25.3 | Right palm | 18.45 | 1.5 | 3.0 |
| 60 | Male | Left Transradial | 27.2 | Left residual limb, palmar reinnervation | 12.03 | 6.1 | 12.2 |

## D. Experimental Design

We quantified the just-noticeable differences using a two-alternative forced-choice paradigm. The participants were presented with two one-second stimulus trains separated by a one-second inter-stimulus interval. The participants were asked to respond to which of the two stimulus trains was more intense.

The participants were allowed as much time as necessary to respond to the two-alternative forced-choice questions. All participants noted changes in the modality of the percepts due to variations in pulse frequency (e.g., lower frequencies reported as tapping and higher frequencies reported as vibration). Participants were instructed to ignore any changes in quality, duration or location of the sensations and to focus solely on the intensity or magnitude of the sensation. Tactile stimuli, regardless of difference in modality or quality, can all be judged on a single intensive continuum [19].

## E. Experimental Parameters

The two stimulus trains were delivered at the same amplitude and pulse-width, but varied in pulse frequency. On each trial, one of the two stimulus trains served as a reference frequency and had fixed frequency (50 or 100 Hz) throughout the experiment. The second stimulus train served as a test frequency that ranged from 25 to 175% of the reference frequency for the 50-Hz and 100-Hz references for intraneural stimulation, from 25 to 175% for the 50-Hz reference frequency for electrocutaneous stimulation, and from 50 to 200% for the 100-Hz reference frequency for electrocutaneous stimulation. For a given experiment, nine different test frequencies were explored and the nine test frequencies were identical to those used in [20]. For the 100-Hz reference electrocutaneous experiments, a tenth test frequency was added at 200% for intact participants or at 250% for the amputee. The order in which the test frequency and reference frequency appeared in a given trial was randomized. A total of 180 or 200 trials were performed for a given reference frequency (20 trials for each of the nine or ten test frequencies) in a randomized order.

## F. Just-Noticeable Difference and Weber Fraction

Discrimination data at the nine or ten test frequencies were fit with cumulative normal distributions to obtain psychometric functions. The just-noticeable difference (JND) was estimated as the change in pulse frequency that the participants' could identify correctly 75% of the time. Each function provided two JNDs (one for decreases and one for increases in pulse frequency) which were averaged. To compare discriminability independent of the reference frequency, we computed the Weber fraction, which is defined as the JND divided by the reference frequency.

## G. Statistical Analyses

All data were screened for normality prior to statistical analyses. A two-sample paired t-test was used to compare the Weber fractions at 50 Hz and 100 Hz across all participants and all stimulation types. Grubbs test for outliers was used to determine if the amputee's intraneural or electrocutaneous Weber fractions were statistical outliers relative to the intact participants' electrocutaneous Weber fractions.

## III. RESULTS

### A. Intensity discrimination of electrocutaneous pulse frequency was consistent between intact participants and a transradial amputee

To explore the maximum number of sensory gradations possible with electrocutaneous stimulation, we had seven intact participants and one transradial amputee discriminate perceived intensity as a function of electrocutaneous pulse frequency. The detection threshold was considerably higher for the amputee's reinnervated skin than for the palm of intact individuals (Table 1). However, the psychometric functions for intensity discrimination of electrocutaneous pulse frequency were similar between the intact participants and the amputee (Fig. 1A). The JND for electrocutaneous pulse frequency in intact individuals was 6.71 ± 2.15 Hz and 34.48 ± 13.32 Hz at the 50-Hz and 100-Hz references, respectively. The JND for electrocutaneous pulse frequency in the amputee was 5.06 Hz and 29.83 Hz with the 50-Hz and 100-Hz references, respectively.

### B. Intensity discrimination of intraneural pulse frequency was similar to that of electrocutaneous pulse frequency

We also had the amputee discriminate the frequency of electrical pulses delivered intraneurally. We found that the psychometric function for intraneural pulse frequency was similar to that for electrocutaneous pulse frequency in intact participants (Fig. 1B) and for the amputee (Fig. 1C). The JND for intraneural pulse frequency in the amputee was 5.16 Hz and 25.03 Hz at the 50-Hz and 100-Hz references, respectively.

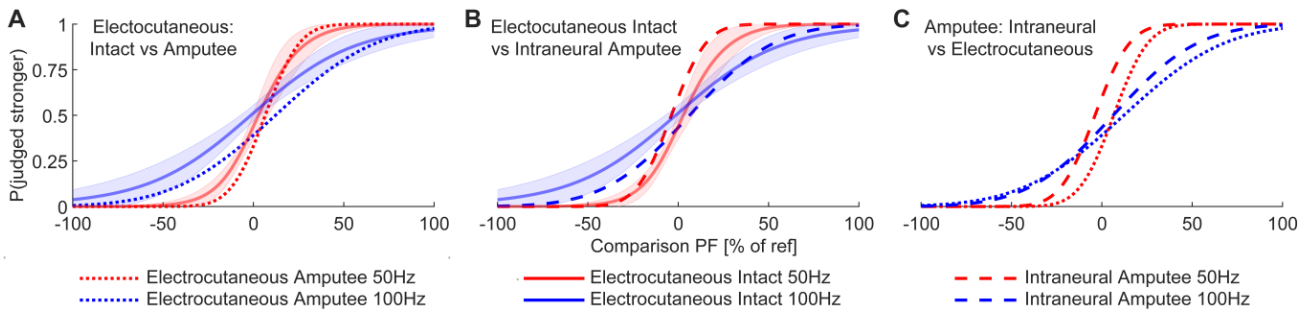

Figure 1: Psychometric functions relating intensity discrimination performance to changes in pulse frequency. Discrimination performance is given as the percentage of test stimuli identified stronger. Pulse frequency (PF) is reported as a percentage of the reference pulse frequency (50-Hz reference in red, 100-Hz reference in blue). **A)** Intensity discrimination of electrocutaneous stimulation was similar between seven intact participants (solid lines, mean ± STD) and one transradial amputee (dotted lines). **B)** Intensity discrimination of intraneural microstimulation in one transradial amputee (dashed lines) was also similar to intensity discrimination for electrocutaneous stimulation with intact participants. **C)** For the same transradial amputee, intensity discrimination was comparable between intraneural microstimulation and electrocutaneous stimulation.

*C. Weber fractions at the 50-Hz reference pulse frequency were significantly lower than Weber fractions at the 100-Hz reference pulse frequency*

To compare sensitivity across reference frequencies, we calculated the Weber fraction, defined as the JND divided by the reference frequency (Table 2). We found sensitivity as gauged by the Weber fraction to be similar for the intraneural and electrocutaneous stimulation of the amputee and electrocutaneous stimulation of the intact participants ($p > 0.05$, Grubbs test). However, sensitivity with the 50-Hz reference was significantly better (Weber fractions were lower) than with the 100-Hz reference ($p < 0.001$, paired t-test; Fig. 2).

Table 2. Weber Fractions

| Reference Frequency | Electrocutaneous Intact | Electrocutaneous Amputee | Intraneural Amputee |
|---|---|---|---|
| 50 Hz | 0.13 ± 0.04 | 0.10 | 0.10 |
| 100 Hz | 0.34 ± 0.13 | 0.30 | 0.25 |

## IV. DISCUSSION

Electrocutaneous stimulation of reinnervated skin on the residual limb may constitute a non-invasive approach to provide intuitive sensory feedback. We demonstrate that the ability to discriminate changes in pulse frequency is comparable between a transradial amputee and seven intact individuals. We also show that the discriminability of pulse frequency is comparable between electrocutaneous and intraneural stimulation. Lastly, for both electrocutaneous and intraneural stimulation, we demonstrate that pulse frequency is nearly three times more discriminable at lower frequencies than it is at higher frequencies.

Weber fractions for epineural stimulation have been reported as 0.25 at a 15-Hz reference [21], 0.33 at a 50-Hz reference, and 0.30 at a 100-Hz reference [20]. We obtained a Weber fraction of 0.1 with the 50-Hz reference, much lower than the previously measured value. Our Weber fraction with the 100-Hz reference, however, are consistent with previously measured values using epineural stimulation via chronically implanted nerve cuff electrodes.

One possible explanation is that at lower frequencies, there are supplementary cues to help distinguish pulse frequency. For example, at lower frequencies individual pulses can be felt, and this could allow for discrimination based on timing rather than just intensity. However, the methodology, reference frequencies, and test frequencies adopted here are similar to those in previously studies [20]. Furthermore, a higher Weber fraction while using cuff electrodes was also found when participants were verbally instructed to identify higher frequency, not necessarily intensity [21].

Selectivity is a major distinction between intraneural and epineural stimulation. Epineural stimulation activates a larger area of the nerve and elicits sensations projected to larger swaths of skin than does intraneural stimulation [13], [15], [21]. For the present study, the projection fields for both intraneural and electrocutaneous stimulation were limited to a 1-cm to 3-cm diameter circle on the palm between the index finger and thumb. Smaller projection fields and/or a more selective activation of afferent fibers may have contributed to a smaller Weber fraction for electrocutaneous and intraneural stimulation. Larger projection fields from epineural stimulation may also coincide with increased paresthesia that could mask subtle percepts.

The present study highlights that functional discrimination of pulse frequency is comparable between intraneural microstimulation via chronically implanted intrafascicular electrodes arrays (USEAs) and non-invasive electrocutaneous stimulation. The electrocutaneous experiments were performed 2.5 years after the USEAs were explanted. However, the participant remarked, "it feels as close as we were able to get before [with intraneural microstimulation]," in response to the electrocutaneous sensory feedback on both the residual limb and intact contralateral hand. Electrocutaneous stimulation may provide a unique opportunity to assess the performance of intraneural microstimulation using electrocutaneous stimulation in intact participants. Future work should investigate the ability to improve electrocutaneous stimulation using biomimetic stimulation patterns, similar to what was recently demonstrated with intraneural microstimulation in amputees [13], [22].

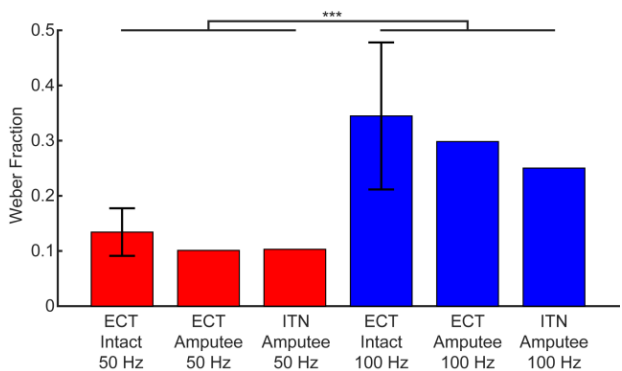

Figure 2: Weber fractions for intensity discrimination as a function of pulse frequency. Weber fractions at 50-Hz references (red) were significantly lower than the Weber fractions at 100-Hz references (blue). Weber fractions for intraneural (ITN) and electrocutaneous (ECT) stimulation with a transradial amputee were not statistical outliers from the Weber fractions for electrocutaneous stimulation with seven intact participants. Data show mean ± S.E.M. *** $p < 0.001$, two-sample paired t-test.

## V. Conclusion

Our sense of touch is critical part of who we are and what we can do. Here we demonstrate that two forms of artificial sensory feedback—invasive intraneural microstimulation and non-invasive electrocutaneous stimulation—are comparable in their ability to convey the magnitude of tactile stimuli. In addition, we show that intraneural and electrocutaneous stimulation both offer improved discriminability at lower frequencies relative to epineural stimulation. These results can help guide the development of sensorized bionic arms with an artificial sense of touch.

## Acknowledgment

This work was funded by: DARPA, BTO, Hand Proprioception and Touch Interfaces program, Space and Naval Warfare Systems Center, Pacific, Contract No. N66001-15-C-4017; NSF Award No. ECCS-1533649; NSF GRFP Award No. 1747505; and The University of Utah Undergraduate Research Opportunities Program.

## Author Contributions

JAG developed sensory software, designed experiments, performed intraneural experiments, performed data analysis and wrote manuscript. MRB designed and oversaw development of the electrocutaneous stimulator and electrocutaneous experiments. PCC performed electrocutaneous experiments. GKC designed and built the electrocutaneous stimulator. SJB helped design experiments. GAC oversaw all aspects of the research. All authors contributed to revising the manuscript.